\documentclass[aps,prl,twocolumn,epsfig]{revtex4}
\usepackage{graphicx}
\begin{document}
\title{Droplet detachment and bead formation in
visco-elastic fluids}
\author{C.~Wagner$^{1,*}$, Y.~Amarouchene$^{2,3}$, D.~Bonn$^{3,4}$
and J.~Eggers$^{5}$}
\affiliation{$^1$ Experimentalphysik,  Universit\"at des
Saarlandes, Postfach 151150, 66041  Saarbr\"ucken, Germany  \\
$^2$ Centre de Physique MolÈculaire Optique et Hertzienne,
Universit\'{e} Bordeaux 1 (UMR 5798), 351 cours de la liberation,
33405 Talence, France \\ $^3$ Laboratoire de  Physique
Statistique, UMR CNRS 8550, Ecole Normale Sup\'{e}rieure, 24  rue
Lhomond, 75231 Paris Cedex 05, France \\  $^4$ van der
Waals-Zeeman Institute, University of Amsterdam, Valckenierstraat
65, 1018 XE Amsterdam, The Netherlands \\ $^5$ School of
Mathematics, University of Bristol, University Walk, Bristol BS8
1TW United Kingdom    $^*$ c.wagner at mx.uni-saarland.de}

\begin{abstract}
The presence of a very small amount of high molecular weight
polymer significantly delays the pinch-off singularity of a drop
of water falling from a faucet, and leads to the formation of a
long-lived cylindrical filament. In this paper we present
experiments, numerical simulations, and theory
which examines the pinch-off process
in the presence of polymers.
The numerical simulations are found to be in excellent agreement
with experiment. As a test case, we establish the
conditions under which a small bead remains on
the filament;  we find that this is
due to the asymmetry induced by the self-similar pinch-off of the droplet.
\end{abstract}
\maketitle
The pinch-off of liquid drops has attracted
considerable attention in recent years
\cite{Eggers97}, not the least owing to its
enormous technological applications in
biotechnology, micro-scale manufacturing, and
spray technology  \cite{Wallace01, Basaran02,Villermaux04}.
More recently, the breakup of
droplets of visco-elastic liquids has been
studied both experimentally and theoretically.
The interest of considering visco-elasticity is
both fundamental and applied. From the
fundamental side, the addition of minute amounts
of polymers has been shown to inhibit the
finite-time singularity that happens at breakup.
From the applied side, in many applications, such as
fire fighting, inkjet printing, or
pesticide deposition on plant leaves, complex
fluids have been used to control or modify drop sizes.

For drop detachment in Newtonian fluids, the
dynamics is governed by the mathematical properties of the
similarity solutions which describe
the thinning of the fluid neck. Most importantly
for applications, these similarity solutions are
always {\it independent of initial conditions},
and hence it is very difficult to control drop
breakup by an external manipulation of
parameters. Therefore, the opportunities offered
by complex fluids are tremendously important.

\begin{figure}
\includegraphics[width=0.7\hsize]{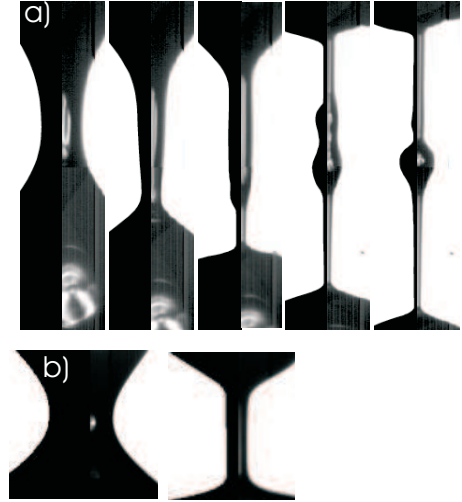}  \caption{The
droplet detachment process of an aqueous 100ppm PEO solution at
different time steps $t_c-t$ relative to the time $t_c$ at which
the filament is formed. The left half of the pictures are plots
from numerical simulations, the right half are experimental
photographs : a): nozzle radius R = 1.5 mm (pictures are 1.2 x 9
mm) $t_c-t$= 6, 2, 0, -3, -5 ms; b: R = 0.4 mm (pictures are 0.9 x
0.9mm) $t_c-t$= 1, 0 ms. The model parameters used for the
numerical simulation were $\eta_p=3.7\times 10^{-4} Pa s$ for the
polymeric contribution to the viscosity, a polymer timescale of
$\lambda=1.2\times 10^{-2}$, and an elasticity parameter of
$b=2.5\times 10^4$. The solvent viscosity (water) is
$\eta_s=1\times10^{-3} Pa s$ and the surface tension is
$\gamma=6\times10^{-2} N / m$.} \label{fig1}
\end{figure}

In this Letter, we consider the detachment
process in the simplest possible visco-elastic
case, that of a dilute aqueous solution of
flexible polymers. The detachment is studied
experimentally for a well-characterized polymer.
This allows us to use a model for the elasticity
due to the introduction of the polymers, and
perform numerical simulations of the
detachment process. The good agreement between
experiment and numerics  permits to study
the detachment process in great detail. As a test
case, we investigate the condition under which the
so-called ``beads on a string'' structure forms \cite{GYPS69},
shown in Fig.\ref{fig1}a. This intriguing structure is
ubiquitous in complex fluids, having been observed for
solutions of both flexible and rigid polymers in
both low-and high viscosity solvents, and various
semi-dilute and concentrated wormlike micellar
surfactant solutions. However, depending on the
experimental conditions, beads may or may not
form.

For the experiments, we focus on the particular case of a drop
falling from an orifice (cf. Fig.\ref{fig1}), although the theoretical
arguments we develop here are more general, and
also apply for instance to filament stretching or
filament thinning rheometry.  Experiments
were performed with aqueous solutions of
polyethylenoxide (POE) with a molecular weight of $4 \times 10^6$
amu, in concentrations of $5$ to $2000$ ppm. Nozzle radii $R$
ranged from $R = 0.25 mm$ to $R = 5 mm$ and the droplets were
generated using a syringe pump in a quasi static mode. Pictures
were taken with a high speed camera (1000 frames/s).

We find that the initial stages of the thinning of the liquid column are well
described by Rayleigh's linear theory \cite{Eggers97}, which
predicts an exponential growth of the most unstable mode. Thus a
growing trough forms on the liquid bridge that separates
the falling drop from the tap, whose shape is fairly symmetric
around the minimum bridge thickness, as seen in the first panel of
Fig.\ref{fig1}a and b. In the second panel of Fig.\ref{fig1}a
the neck shape has already turned asymmetric, while for the
smaller pipette radius the shape turns directly to a uniform
cylinder. For the asymmetric case, the remaining frames of
Fig.\ref{fig1}a show how the asymmetric neck structure turns
into a secondary bead.

In order to model the detachment, we use the
FENE-P (Finite Extensibility Non-linear
Elasticity) model \cite{Bird87}, a simple polymer model which
treats the polymers as Hookean springs of finite
extensibility. The springs can be stretched by
the flow, leading to elastic stresses due to the
entropic elasticity of the polymer chains. To
further simplify the description,
we use a 1d long-wavelength theory for the fluid
motion \cite{Eggers97}, leading to a coupled set of equations
for the local radius h(z,t) of the fluid column, the mean fluid
velocity v(z,t), and the radial and axial components of the
polymeric stress $\sigma_r(z,t), \sigma_z(z,t)$. For the present
problem, only the component in the axial direction $\sigma_z(z,t)$
is of interest, because this is the principal direction of
stretching.

The model parameters used in the simulation were taken from the
rheological measurements of \cite{Lindner03}. To obtain an optimal
description, we used a polymer relaxation time $\lambda =
1.2\times 10^{-2}s$, about 4 times as long as suggested
by \cite{Lindner03}, from steady-state
rheological measurements. Our comparison
theory/experiment can be seen as a method of
determining
the timescale relevant for polymer-flow interactions in a highly
non-equilibrium situation. Given the considerable simplification
inherent in a single-time-scale approach, we believe the
correspondence in time scales between the rheological method and
the value used here to be quite reasonable.

In Fig. \ref{fig1} experiments are compared to numerical
simulations, finding
excellent agreement for the two different pipette radii was obtained
by adjusting a single parameter, the polymer relaxation time
$\lambda$. For the larger pipette radius of $1.5mm$ a secondary
bead is formed (cf. Fig.\ref{fig1} a)), for the smaller radius of
$0.4mm$ the elastic thread is completely uniform.

\begin{figure}
\includegraphics [width=0.8\hsize] {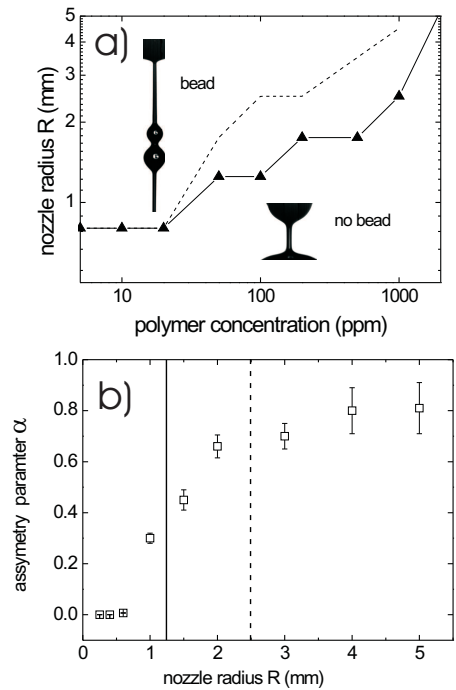}
\caption{a)The phase boundary for bead formation,
obtained by analyzing experimental runs at polymer concentrations
of $5, 10, 20, 50, 100, 200, 500, 1000$ and $2000$ ppm and nozzle
radii $R = 0.25, 0.75, 1, 1.5, 2, 3, 4, 5 mm$(triangles).
Between the dashed and the full lines a bead is formed, but
subsequently disappears.
b): The asymmetry parameter $\alpha$ for the the 100 ppm runs
measured at $t_c-t=1ms$. The vertical lines correspond to the
transition lines in a).} \label{fig2}
\end{figure}

Fig. \ref{fig2}a shows a phase diagram for the existence of a
bead. We find that for low polymer concentrations and large nozzle
radius a bead is observed, while in the opposite corner, below the
full line, no bead is formed. In the narrow strip between the
dashed and the full lines a bead is formed, but is so small that
it gets stretched out by elastic stresses and eventually
disappears.

From the data shown in Fig.1, it appears that the
transition to the asymmetric neck
plays an important role for the formation of  the
secondary bead structure. We quantify the
asymmetry using the parameter $\alpha$
\cite{Rhotert01}:
\begin{equation}
\alpha=\frac {\int{h\left(z_{min}+z' \right)-h\left(z_{min}-z'
\right) dz'}}{\int h\left(z_{min}+z' \right) +h\left(z_{min}-z'
\right)-2h_{min} dz'} , \label{eqn1}
\end{equation}
whose value is zero for a symmetric and one for a staircase
profile. The boundaries of integration were chosen at
$h\left(\pm z'\right)=2 h_{min}$.
In Fig.\ref{fig2}b we show a cut
through the phase diagram at constant polymer
concentration. Indeed, we find that the
asymmetry parameter $\alpha$ just before onset of the filament
undergoes a rather abrupt transition from zero to
a nonzero value, exactly at the point
corresponding to the boundary between the bead and no-bead regions of
Fig.\ref{fig2}a
in the phase diagram. Similar cuts at constant nozzle radius show the
same behavior.

\begin{figure}
\includegraphics [width=0.8\hsize]  {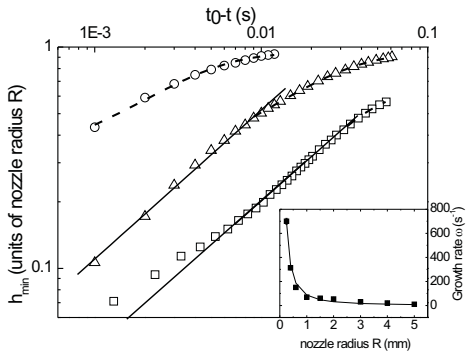}
\caption{The minimum neck radius versus time for $R = 0.4 mm$
(circles), $R = 1 mm$ (triangles), and $R = 4 mm$ (squares),
at a concentration of $100 ppm$, where $t_0$ is the
extrapolated time at which pinch off would take place for a
Newtonian liquid. The straight lines are power-law fits
yielding an exponent of $0.66 \pm 0.01$.
The dashed lines are exponential fits to the disturbance amplitude
$1-h_{min}/R$. The squares in the inset are the corresponding
growth rates, the line is the theoretical prediction for the most
unstable Rayleigh mode of an inviscid fluid. } \label{fig3}
\end{figure}

The transition from the symmetric to the asymmetric shape can be
understood by considering the Newtonian motion alone, since the
onset of elastic effects is very abrupt for low-viscosity
solvents. We confirm this observation by noting that the motion of
the minimum neck radius up to just {\it before} the onset of
elastic effects is well described by Newtonian theory, see also
\cite{Amarouchene01}. For the regime before the
formation of the filament, at early times, the
motion is consistent
with Rayleigh's theory of exponential growth.
More quantitatively, this leads to the growth of
the disturbance of
the liquid column with a rate  $\omega_R = \left(0.118
\gamma/(\rho R^3)\right)^{1/2}$, with $\gamma$ the surface tension
and $\rho$ the liquid density. Corresponding fits to $h_{min}$ are
shown in Fig. \ref{fig3} (dashed lines), and compared to theory in
the inset. Fore later times, as the bridge becomes more strongly deformed, the
well-known similarity solution \cite{Day98} $h_{min} = 0.7
(\gamma(t_0-t)^2/\rho)^{1/3}$ for the detachment
of a low-viscosity droplet takes over. Here $t_0$
is the extrapolated singularity time, at which
the droplet would pinch off in the absence of polymer. This
power law corresponds to the straight lines in
Fig.\ref{fig3}. We see in Fig.\ref{fig3} that
the power-law behavior is not observed for the
smallest nozzle radius: we have not yet transited
into the similarity solution regime before the
elastic forces due to the polymer intervene and
lead to the formation of the filament.

As the Rayleigh theory is a linear stability
argument, and the similarity solution is only
valid close to the pinch-off, there is no theory
for the transition between the two regimes. However experimental
observations show that the transition from an asymmetric to a
symmetric shape  occurs at a fixed value of $h_{min}/R = 0.17
\pm0.01$, corresponding to a sufficiently strong deformation of
the liquid bridge.

Bead formation can now be understood from the interplay between
inviscid dynamics and elastic effects: in regions where significant
polymer stretching occurs a uniform thread is formed. If elongation
rates during the initial Rayleigh thinning are sufficiently high
for the polymers to stretch, the symmetric Rayleigh solution
transforms directly in a uniform thread, as shown in Fig. \ref{fig1}b.
For larger values of $R$, however, the elongation rates are smaller
and polymer stretching sets in later, such that the asymmetric
similarity solution develops without being
influenced by the presence of the polymers.

From Fig.\ref{fig1}a it is evident that the bead starts to form at
the bottom of the filament, near the droplet, where the fluid neck
is thinnest. Since the elongation rate becomes very large in this
region, a filament forms in a very localized fashion. Meanwhile,
the rest of the fluid neck continues to evolve as if the fluid were
inviscid, eventually pinching at the top of the neck. As a result,
another filament forms, isolating a bead in the middle (cf. Fig. \ref{fig1}a).
Thus the bead is the exact analogue of the so-called ``satellite drop''
\cite{Eggers97},
which is always formed between the main drop and the faucet in the
case of a Newtonian fluid. However, the polymer prevents pinch-off,
and a thread remains, connecting the ``satellite drop'' to both the
main drop and the faucet.

If the growth rate $\omega_R$ of the Rayleigh
instability exceeds $\lambda^{-1}$, such that  polymer stretching
takes place from the start, no bead is formed. Thus
$R_{crit}=(0.118\lambda^2\gamma/\rho)^{1/3}\approx 1mm$
is the critical radius above which a bead is formed,
in good agreement with experiment for low polymer concentrations.
The influence of polymer
concentration, is relatively minor, in agreement
with this argument: the critical radius changes
only by  a factor of $two$ while the
concentration varies by two orders
of magnitude. The slight increase of $R_{crit}$
with concentration reflects an increase in the relaxation time
$\lambda$, perhaps indicating overlap \cite{TMC04}.

The growth of elastic stresses due to the stretching of the polymer,
can be calculated as follows. The last stages of the
inviscid motion are very fast, so the polymer does not have
time to relax. In such a case the axial polymer stress is given by
the total {\it deformation}, leading to $\sigma_z=\sigma_0/h^4$
\cite{Entov84,Clasen03}, in excellent agreement with the results
of our numerical simulations. The constant
$\sigma_0=\eta_p h_0^4/(\rho\lambda)$ \cite{Clasen03} is determined by the
neck radius $h_0$ at which the timescale of the flow has become
shorter than the polymer timescale $\lambda$, where $\nu_p$ is the
polymeric contribution to the viscosity.

For simplicity let us focus on the smallest pipette radii, for which
we can estimate that $h_0\approx R$;
equating elastic and surface tension forces, we infer that  threads start
to form when the minimum radius is
$h_{thread}/R\approx (\eta_p R/(\lambda\gamma))^{1/3}$. This
means the total deformation the polymers undergo before
a thread forms is proportional to $\lambda^{-4/3}$, where
$\lambda$ is proportional to the solvent viscosity, and thus
relatively small for water. Our experiments
show that the thinning of the filament is exponential, i.e. the
elongational rate $\dot{\epsilon}_0$, is constant.
Exponential thinning of filaments is generally believed
\cite{Anna00, Amarouchene01, Clasen03} to result from a balance of surface
tension and elastic forces, concomitant with exponential
stretching of polymers.

The present simulations however do {\it not} agree with this simple
picture: by the time filaments are formed, polymer extension is so
strong that non-linear effects become important within the
present FENE-P modeling. We have chosen the finite extensibility
parameter $b=2.5\times 10^{4}$, in agreement with \cite{Lindner03}
and obtain reasonable agreement between the observed
thinning rates in the simulation and the experiments. However,
in the simulations the thinning rate  only  remains constant for
less than a decade, in disagreement with the experimental
observations.
Some experimental confirmation of non-linear effects in the
constitutive equations is provided for by our finding of a weak
dependence of $\dot{\epsilon}_0$ on the nozzle radius. However,
serious questions remain as for the applicability of the FENE
model to strongly nonlinear phenomena in dilute polymer solutions
(\cite{Wagner04}).

\begin{figure}
\includegraphics [width=0.5\hsize] {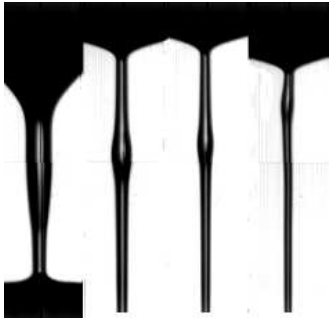}
\caption{The migration of the primary bead for the 50ppm, $R = 1.5
mm$ run (pictures are 1 x 4 mm). The profiles are shown in time
steps of $2 ms$. The bead is pushed upward against the direction
of gravity. } \label{fig4}
\end{figure}

After the discussion of secondary bead formation it is worth
examining the subsequent evolution of beads, which can be quite
varied. Li and Fontelos \cite{Li03}
predict several scenarios of which we found at least three in our
experiments: The disappearance of a bead due to stretching, bead
migration, and fusion of threads due to differences in the
capillary pressure. Bead stretching (between the dashed and full lines
in Fig. \ref{fig2}) occurs if the bead is small and the pressure
difference between the bead and the surrounding filament is small.
Bead migration usually follows gravity, but smaller beads may be
driven upward by pressure forces (cf. Fig. \ref{fig4}), where they
fuse with the upper reservoir. Small beads also get sucked into
larger ones (cf. Fig. \ref{fig1}a), because their internal pressure is
higher.

In conclusion we have measured and described the
detachment of a drop of a visco-elastic liquid,
and determined the conditions for the formation
of a secondary bead in a the detachment process. Our numerical
simulations describe the drop formation process in very favorable agreement
with the experiments, and allow us explain
the qualitative features of the 'phase diagram'
for the formation of secondary beads. However,
for later times strong non-linear effects become
important, and the agreement between experiment
and simulation is less good.

\end{document}